# Self-assembly near Ground State: Randomly Pack Granular Spheres and Cubes into Crystal

Reza Amirifar[1], Kejun Dong[1,*] and Aibing Yu[2,3,†]

[1]*Centre for Infrastructure Engineering, School of Engineering, Western Sydney University, Sydney, Australia*
[2] *ARC Research Hub for Computational Particle Technology, Department of Chemical Engineering, Monash University, Clayton, VIC 3800, Australia*
[3] *Australia Centre for Simulation and Modelling of Particulate Systems, Monash-Southeast University Joint Research Institute, Suzhou, 215123, China*

*Corresponding author's E-mail: Kejun.Dong@westernsydney.edu.au
†Corresponding author's E-mail: Aibing.Yu@monash.au

**Abstract:** Self-assembly of granular particles is of great interest in both applied and basic research. It is commonly observed that when randomly packed into a container, granular particles form disordered structures like glass. As the particles are athermal, the self-assembly of such packings can normally be directed with energy input via vibration or shear. However, here we show that in particular containers, mono-sized spheres and cubes can self-assemble into perfect crystals when randomly dropped in. This is because the favourable microstates for new particles are jammed in the ordered structure by the existing particles and the boundary synergistically. Such a self-assembly method has not been reported in the literature. It indicates that disordered packing structure may result from the conflict between the internal structure and the structure shaped by the boundary. Therefore, to bridge such inconsistency could be a general principle for directing self-assembly for different kinds of particles in emerging areas.

Self-assembly of discrete elements is receiving increasing attention in emerging areas from metamaterials [1] to natural [2] and artificial beings [3]. However, how to direct self-assembly for granular particles, one of the simplest discrete systems, has not been well understood. In fact, the packings of granular particles have been serving as primitive models for different matters and systems for decades [4-7,8-10]. These packings often form disordered structures like glass [11-14]. A typical example is that uniform non-cohesive spheres will form random packings with packing fraction ranging from 0.55 to 0.68 [8,15-18]. Although these disordered packings are much looser than that of the ordered packings of face-centred-cubic (FCC) or hexagonal-close-packed (HCP) structure with packing fraction of 0.74 [19,20], the particles will not self-assemble to the ordered structures unless with energy input via shear [21-23] or vibration [24-28], as granular particles are athermal. The energy input via vibration or shear can agitate particles to rearrange into ordered structures to certain extent. But how to control such process remains unclear [8,18,29], like recently it has been found that different motions of boundary could have different effects on such self-assembly [14,30]. Nevertheless, if there are no pre-set template particles [31-33] in a container, none has observed that particles can *"self"* assemble when randomly packed into the container.

Here we show an exception in Fig. 1. In an inverted tetrahedron, which is a half-square-pyramid (HSP), if mono-sized spheres are randomly dropped one by one or at a small feed rate, they automatically form a single FCC crystal. No energy input or any other interventions like template particles are needed. We did experiments with ping-pong balls, tennis balls and steel ball bearings. The movies can be found in the Appendix. We also used first principles simulations to confirm that this self-assembly happens for mono-sized spheres of different material properties. In the simulations, particles were fed batch by batch from the top of the container with their horizontal positions randomized. The number of particles per batch is denoted as $N_B$ and the interval time between two batches is denoted as $T_B$. Perfect crystal was generated by using $N_B$=1 and $T_B$ = 0.5 secs, while if $N_B$>1, some point defects could be identified and their percentage increased with increasing $N_B$. But the defects were less than 0.2% for $N_B$=5 in all our simulated cases. We even can mix spherical particles with different material properties but of the same size to achieve such self-assembly. More details can be found in the Appendix.

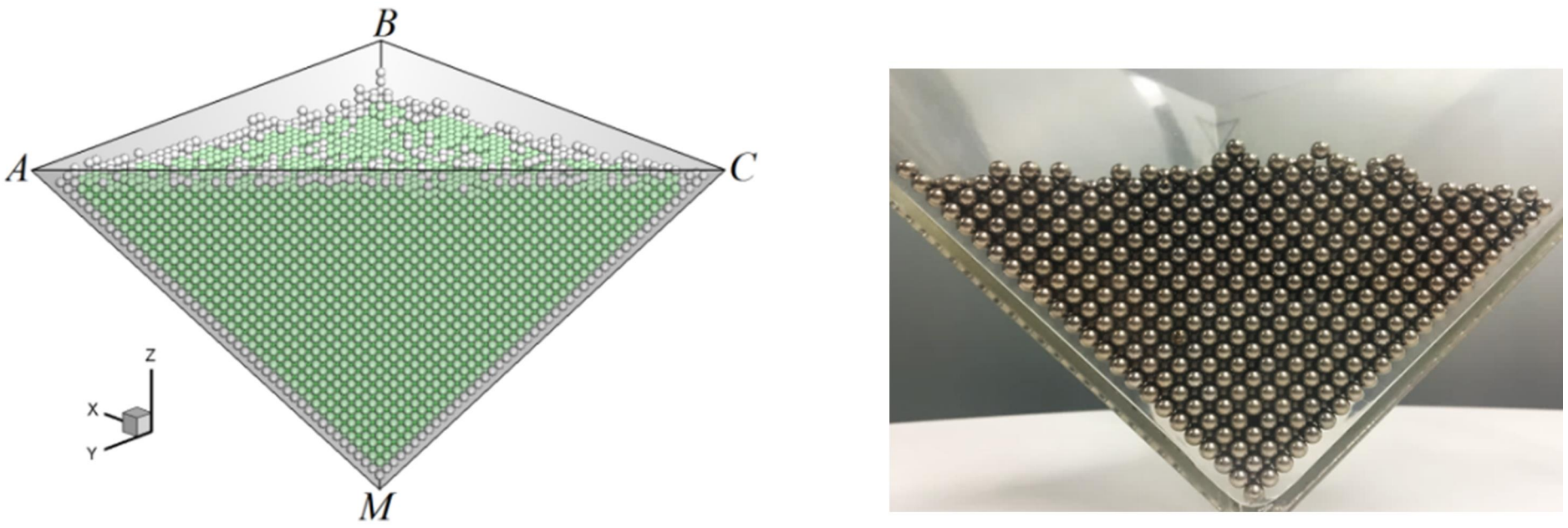

*Simulation* *Experiment*

**FIG. 1.** Mono-sized spheres self-assemble into a perfect FCC when randomly dropped into an inverted half-square-pyramid tetrahedron at a small feed rate, where $AB=BC=AM=BM=CM$ and $AC = \sqrt{2}AB$. $ABC$ is perpendicular to the gravitational direction. Particles are 5mm steel balls. In the simulated packing, the FCC particles identified by Ovito [34] are coloured green.

It should be noted that there are previous studies using different shaped containers or Random-Sequential-Addition (RSA) to form packings, which sometimes can generate ordered packings [24], but none has observed a perfect crystal packing like here with particles just randomly dropped in. There are also studies using a template at the bottom of a container to help form ordered packings for granular particles [32] and colloidal particles [33], but such templates only work for particles with the same size as the template particles, and the formed crystals would be affected by boundary. Here in the HSP tetrahedron, we do not need to put template particles and this container works for different sized particles. Moreover, the ordered structure is actually maintained by the boundary, which will be discussed in detail later. The self-assembly here is either not exactly the same as those found in the Molecular Dynamics (MD) simulated hard sphere systems with the Lennard-Jones (LJ) potential [35-37]. As the MD simulated crystallization is still driven by thermal energy and the LJ potential incurs non-contact interactions between particles, whereas granular spheres are athermal and only present compulsive force when contacted. Nonetheless, even in the LJ simulated hard sphere systems, there are no studies showing such perfect crystallization.

Why here the randomly packed spheres form a crystal rather than the commonly seen disordered packings [17] ? The reason may be more interesting to tell from the process of finding this container. At first, we noticed that boundary has a significant effect on directing ordered structures locally when a packing is vibrated or sheared [24,27,30]. Different shaped containers were used to promote the self-assembly of uniform spheres [24,27,30]. However, in these containers the ordered structures formed from different faces or along different edges may have conflicts in the crystal orientations [30]. We try to design containers that can eliminate such conflicts. Considering a basic container, an inverted tetrahedron corner shown in Fig. 2c, we design the container shape to coordinate ordered structures from low dimension (boundary) to high dimension (system) according to the following steps:

- Particles along an edge should be the 1D JO (Jammed and Ordered) structure, denoted as JO1, in which each particle is contacting with two particles, as shown in Fig. 2a. Note a small gap can be between the edge and the particles as the edge is a part of the tetrahedron.
- Particles on a face should be the 2D JO structure, denoted as JO2. In 2D, the possible JO structures are hexagonal or cubic structures. Therefore, three possible plane angles that can promote the JO2 structures are listed in Fig. 2b, denoted as T1, T2 and T3 respectively. Note particles form the JO1 structure along the edges.
- Since FCC or HCP are the expected 3D structures [19] and FCC is more stable than HCP [38,39], we consider only FCC here. We want the JO1 structures along edges and JO2 structures on faces. Therefore, the three faces of the corner can be selected from T1 to T3 in Fig. 2b. Noticing that the plane angles for T1, T2 and T3 are given, the dihedral angle between two faces can be calculated using these angles. On the other hand, the cubic or hexagonal JO2 structures indicate that they have to be certain crystal faces in a FCC, so the dihedral angle can also be calculated by the face normal vectors. These two calculated angles should agree with each other.

For example, in Fig. 2c, the dihedral angle between faces *AMC* and *BMC* can be calculated by $\alpha = \operatorname{acos}\left(\frac{\cos\widehat{AMB} - \cos\widehat{AMC}\,\cos\widehat{BMC}}{\sin\widehat{AMC}\,\sin\widehat{BMC}}\right)$, where the plane angle ($\widehat{AMB}$, $\widehat{BMC}$ or $\widehat{AMC}$) should be 60, 120 or 90 degrees if the selected face is T1, T2 or T3 respectively. On the other hand, this dihedral angle can also be calculated from the angle between the normal vectors of the two crystal faces in the FCC lattice coordinate system, given by $\alpha' = \operatorname{acos}(\mathbf{n}_{AMC} \cdot \mathbf{n}_{BMC})$ where $\mathbf{n}_{AMC}$ and $\mathbf{n}_{BMC}$ are the normal vectors of *AMC* and *BMC* respectively, and each will be one from $\sqrt{\frac{1}{3}}[\pm1, \pm1, \pm1]$ if the face is T1 or T2, or one from $[\pm1, 0{,}0]$ , $[0, \pm1, 0]$ or $[\,0{,}0, \pm1]$ if the face is T3. An acceptable combination should satisfy that $\alpha = \alpha'$.

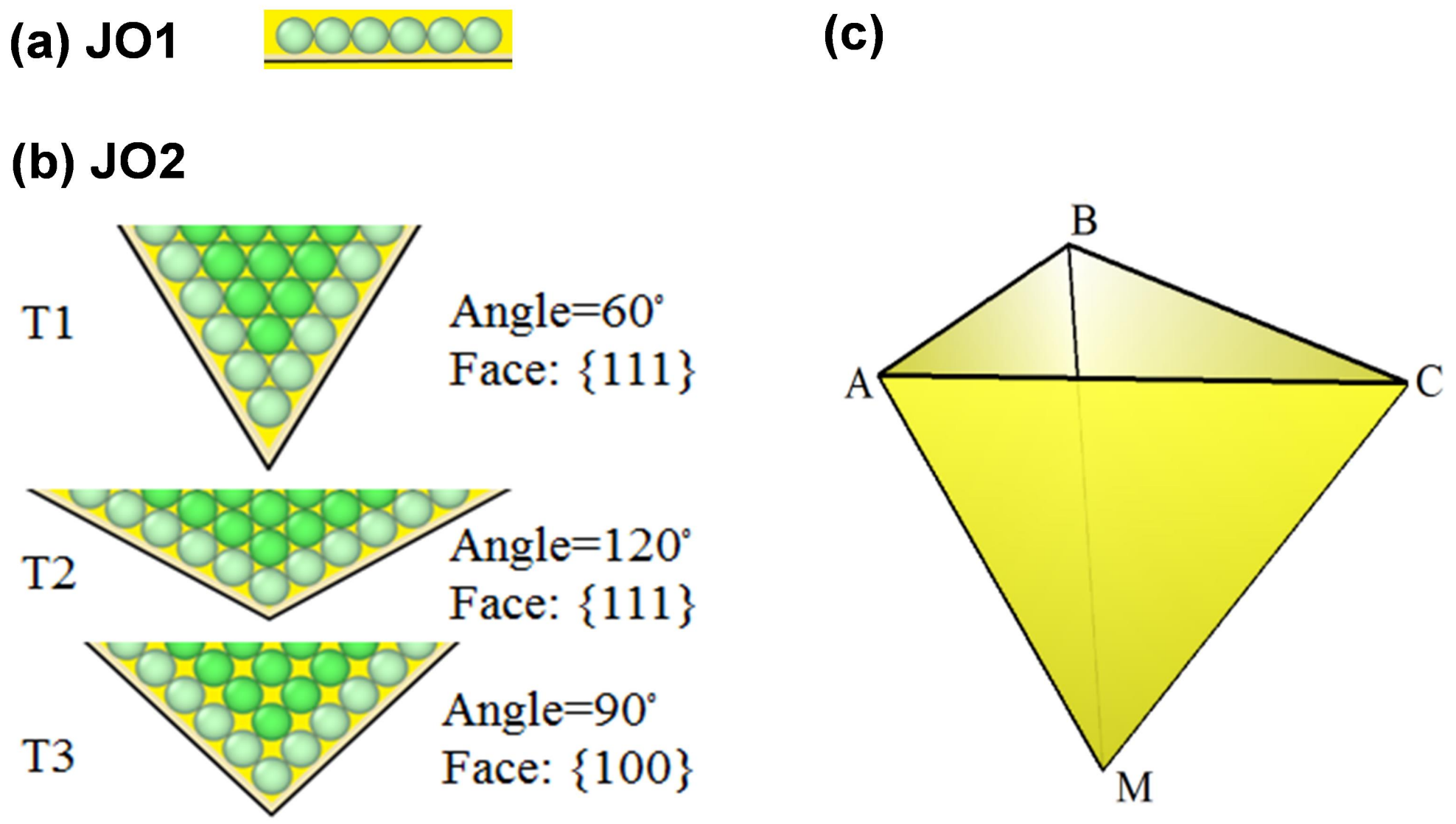

**FIG. 2.** Design a container to promote ordered structures coordinated between the boundary and particles: (a), In 1D, particles compacted along an edge form JO1 structure. (b), In 2D, particles form hexagonal or cubic JO2 structures; three possible plane angles to accommodate both JO1 and JO2 structures are listed as T1, T2 and T3. (c), An inverted tetrahedron to facilitate self-assembly in 3D, the three faces are selected from T1 to T3 in (b).

By applying this rule, we have constructed a series of inverted tetrahedrons, as listed in Table 1, where the first one is the HSP tetrahedron shown in Fig. 1. All of them are very effective in forming a single FCC crystal with simultaneous feeding and vibration, which is a method for self-assembly under vibration introduced in previous studies [26]. However, in previous studies the containers have not been designed to resolve the conflicts of crystal orientations from different boundaries, so a perfect single FCC crystal is difficult to obtain.

**Table 1.** List of inverted tetrahedrons designed based on the JO rule. Perfect FCC packings can be obtained in all containers under 1D vibration governed by $z = A \cdot \sin(2\pi f t)$, where $A$=0.2d and $f$=30Hz. Particles were fed at $N_B$= 10 and $T_B$=0.05s.

| No | Vertexes and Faces<br>(see Fig.2, $a$ is an arbitrary real) | Packing obtained under vibration | Horizontal layer |
|---|---|---|---|
| **1** | $A\left(\frac{-a}{2},\frac{-a}{2},\frac{a}{\sqrt{2}}\right)$<br>$B\left(\frac{a}{2},\frac{-a}{2},\frac{a}{\sqrt{2}}\right)$<br>$C\left(\frac{a}{2},\frac{a}{2},\frac{a}{\sqrt{2}}\right)$<br>AMB: T1; AMC: T3; BMC: T1 | 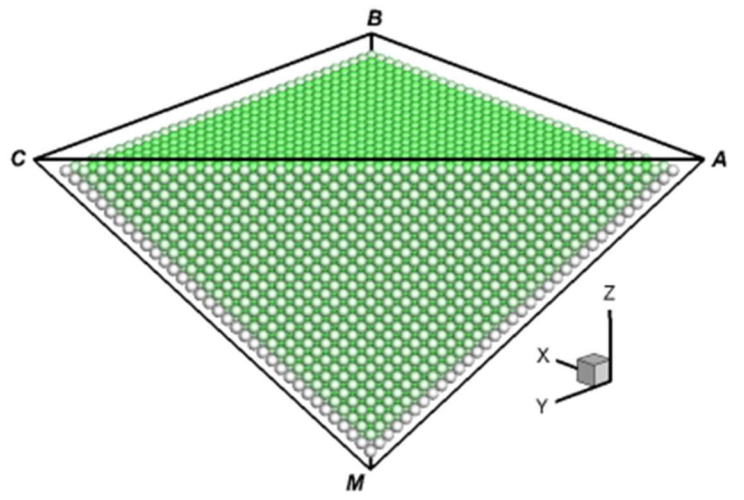 | 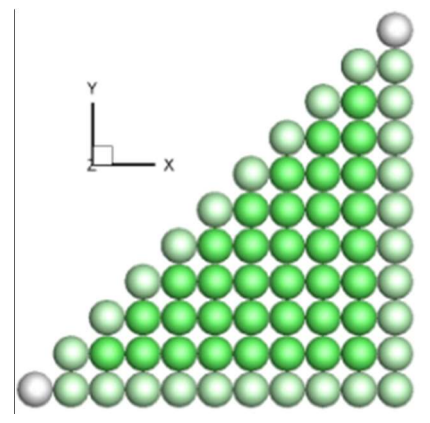<br>{100} Square JO2 |
| **2** | $A\left(0,\frac{a\sqrt{3}}{3},\frac{a\sqrt{2}}{\sqrt{3}}\right)$<br>$B\left(\frac{-a}{2},\frac{-a\sqrt{3}}{6},\frac{a\sqrt{2}}{\sqrt{3}}\right)$<br>$C\left(\frac{a}{2},\frac{-a\sqrt{3}}{6},\frac{a\sqrt{2}}{\sqrt{3}}\right)$<br>AMB: T1; AMC: T1; BMC: T1 | 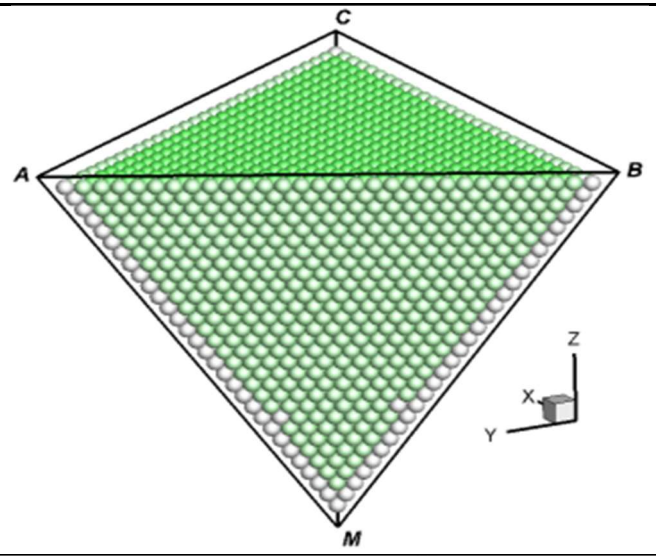 | 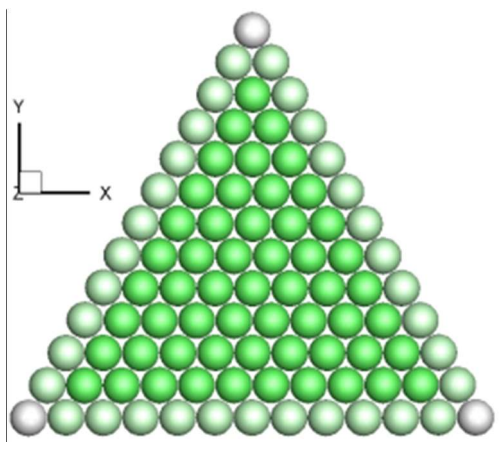<br>{111} Hexagonal JO2 |
| **3** | $A\left(\frac{-3a}{4},\frac{-a}{2},\frac{a\sqrt{3}}{4}\right)$<br>$B\left(\frac{3(2+\sqrt{6})a}{4(6-\sqrt{6})},0,\frac{a\sqrt{3}}{4}\right)$<br>$C\left(\frac{-3a}{4},\frac{a}{2},\frac{a\sqrt{3}}{4}\right)$<br>AMB: T2; AMC: T1; BMC: T2 | 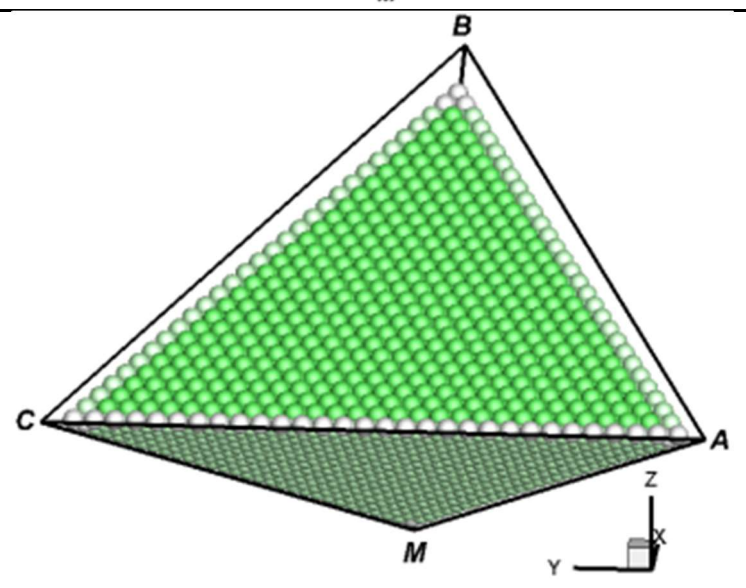 | 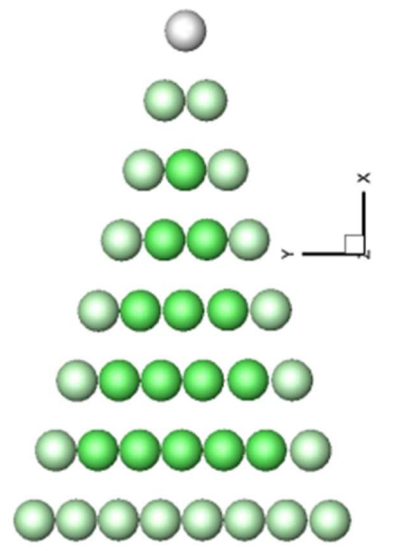 |
| **4** | $A\left(\frac{a\sqrt{6}}{3},\frac{-a\sqrt{3}}{6},\frac{a}{2}\right)$<br>$B\left(0,\frac{a\sqrt{3}}{2},\frac{a}{2}\right)$,<br>$C\left(0,\frac{-a\sqrt{3}}{2},\frac{a}{2}\right)$<br>AMB: T3; AMC: T1; BMC: T2 | 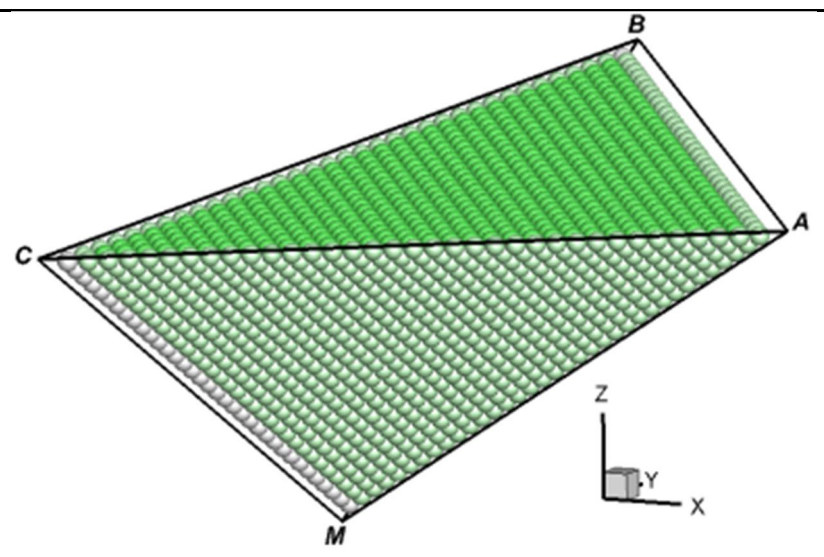 | 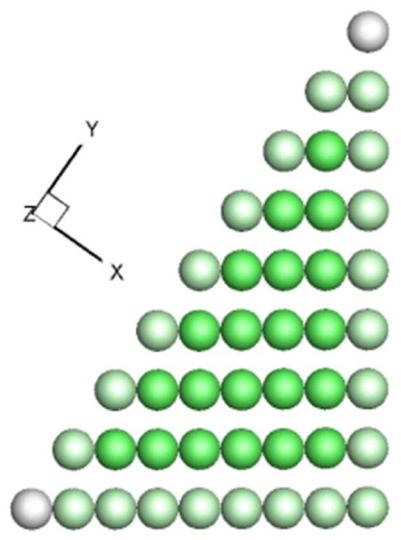 |
| **5** | $A\left(\frac{-a\sqrt{6}}{4},\frac{-a\sqrt{2}}{2},\frac{a\sqrt{2}}{4}\right)$<br>$B\left(\frac{a}{2\sqrt{2}}\text{ctg}\frac{\pi}{12},0,\frac{a\sqrt{2}}{4}\right)$<br>$C\left(\frac{-a\sqrt{6}}{4},\frac{a\sqrt{2}}{2},\frac{a\sqrt{2}}{4}\right)$<br>AMB: T2; AMC: T3; BMC: T2 | 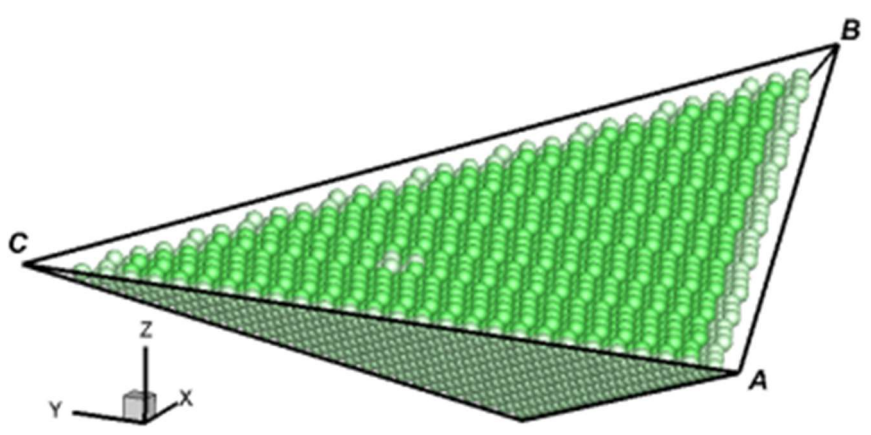 | 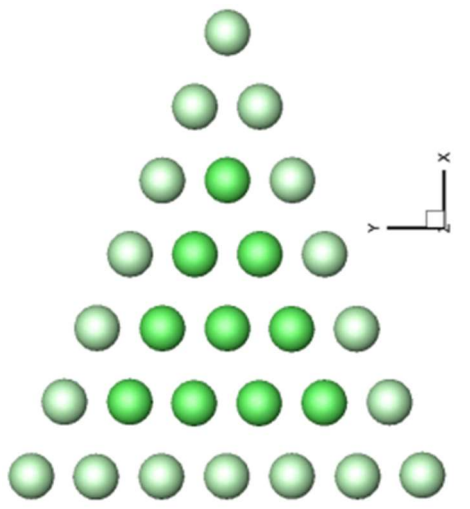 |

Here we focus on the self-assembly without vibration in the HSP tetrahedron. Facilitated by the simulation, we find that the mechanism is hidden at microscopic scale: in this container, a new particle only finds the crystal lattice node positions to be stable. As shown in Fig. 3a and Fig. 3b-1, if the already existing packing consisting of green particles is FCC, the potential (gravitational) energy for a new particle on the top of the bed clearly demonstrates that the local minimum energy points all coincide with the lattice nodes, even at the boundary. Note here the packing is under gravity, so a particle is locally jammed when it is stably supported by the underneath particles and the boundary. For the three white particles in Fig. 3a, particle *A* is jammed along an edge and particle *B* jammed on a face, while particle *C* is jammed only by other particles. Importantly, their jamming positions are in different nodes of the same crystal lattice, so they will not have any conflicts. This complies with our design on the edges and faces. Moreover, the addition of a new particle will keep maintaining the ordered structure by still creating new stable positions only at lattice nodes of the existing crystal, as can be seen from Fig. 3b-2. It is interesting to note that, gravity, which normally jams the particles to be disordered, in this packing protocol, jams them to be ordered.

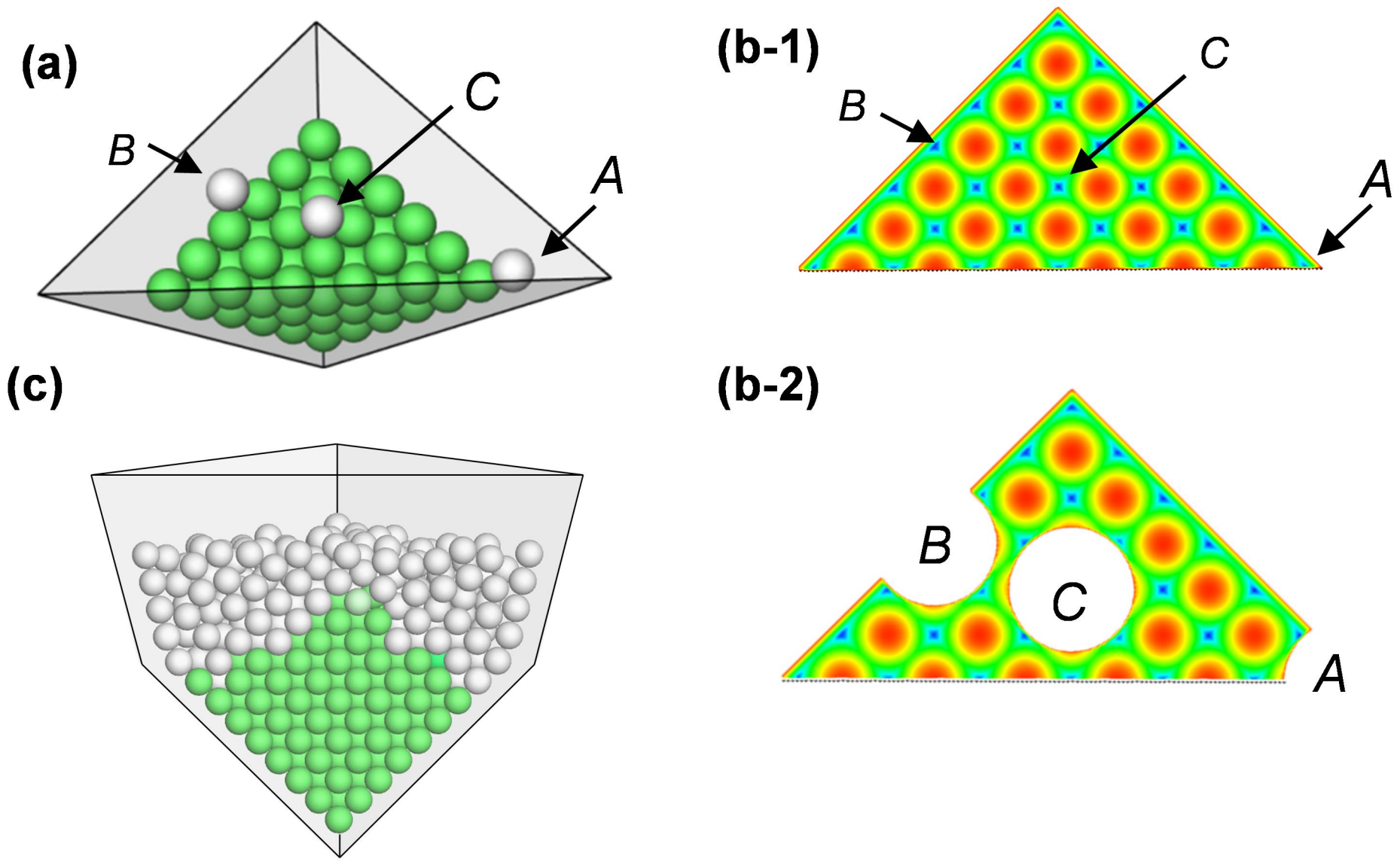

**FIG. 3.** (a), Jammed order in the inverted HSP tetrahedron: all stable positions for the new particles (white) are jammed in lattice positions with existing particles (green) and the boundary. (b), The contour shows the potential energy for a new particle on the top of the substrate as a function of its horizontal position, which decreases from red to blue: b-1, when particles A, B and C are not packed; b-2, when A,B and C are packed. (c), Disorder emerges from the "order-broken" boundary and gradually destroys the crystal.

Compared to the template-induced ordered packings in ref [31] and crystallization of colloidal particles in ref [33], here the container shape plays a critical role. In previous studies, although the template particles played a similar role in directing self-assembly by creating local jamming positions at crystal lattice nodes, the boundary effect was always tried to be avoided, and the crystal was formed in the centre region. For example, Andreea and Arshad [33] observed that disorder nucleated from boundary. But here, the boundary is essentially helping self-assembly, as discussed above. Moreover, if the boundary shape changes, a new particle may find a stable place near the boundary which is not at a crystal lattice node, then disorder emerges from the boundary and gradually affects the whole system. An example is shown in Fig. 3c.

In addition, the HSP tetrahedron has a unique advantage for such crystallization than other containers in Table 1, as the horizontal layers of the FCC packing in it are $\{100\}$ faces, one of the JO2 structures, which are robust in jamming new particles in the FCC lattice nodes. The advantage of the growth of a FCC crystal along $\{100\}$ faces has also been pointed out in previous studies [32,33]. In other tetrahedrons, as shown in Table 1, the horizontal layers of the FCC packings are mostly not JO2 structures, and hence the particles may not be jammed at lattice nodes when randomly fed. Even when the horizontal layers are $\{111\}$ faces (the second one in Table 1), a new particle can be jammed at either HCP or FCC lattice nodes, which leads to the bifurcation of the system. Therefore, only the HSP tetrahedron can direct self-assembly without vibration while self-assembly in other tetrahedrons needs vibration.

Based on the above discussion, if we assume that the self-assembly in the HSP tetrahedron can be decomposed into independent individual processes of each new particle finding a lattice node, the whole self-assembly process can be described by a Markov chain network in the state space. We define the ordered packing states in the HSP tetrahedron as the 3D JO states (denoted as JO3) and identify them in all possible packing states. In Fig. 4, the potential energy $E_g$ as a function of the particle number $N$ is schematically plotted. The maximum $E_g$ can be obtained when the particles are packed along an edge of the HSP tetrahedron, while the minimum $E_g$ can be obtained when the particles form the FCC crystal with a smooth surface [19]. With a given $N$, all possible packings including those disordered ones should be continuously distributed on the vertical line of $N$ between the upper and lower boundaries (the detailed calculations of boundaries can be found in the Appendix). However, the JO3 states can only be some discrete points with split energy levels. Moreover, with a new particle added ($N$ to $N$+1), a JO3($N$) state can only jump to the nearby JO3($N$ + 1) states on the $N$ + 1 line. Then we have: $\sum P_i[\,\mathrm{JO3}(N+1)|\,\mathrm{JO3}(N)] = 1$, where $P_i$ is the probability for a transition from a JO3($N$) to a JO3($N$ + 1) and the summation is over all possible transitions. This demonstrates that the growth path of the JO3 states forms a strict Markov chain network that jams the whole system ordered.

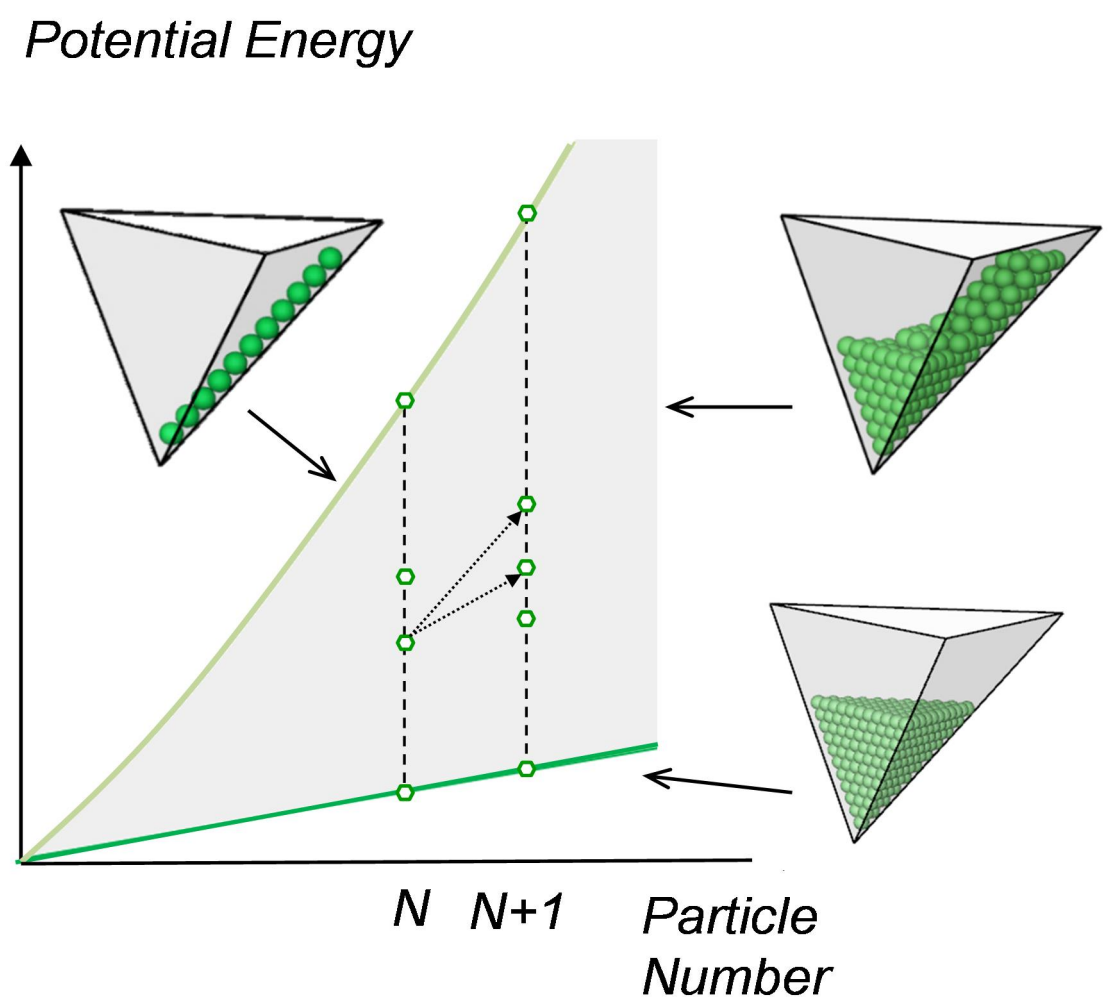

**FIG. 4.** Potential energy vs. particle number for possible packings of mono-sized spheres in the inverted HSP tetrahedron. Packing states are distributed continuously on the corresponding vertical line, but the JO3 states are discrete points on the line, and the state transition paths from $N$ line to $N$+1 line are limited between certain points.

Interestingly, this protocol also works for the self-assembly of cubes. Different methods were used in the literature to direct self-assembly of cubic particles [12,14,40], but all need a kind of energy input. Following the similar idea for the self-assembly of spheres discussed above, in an inverted tetrahedron with three right angles, we observed that the randomly dropped cubic dices can self-assemble into a crystal when randomly fed at a small feed rate, as shown in Fig. 5, no matter who throws the dices. The mechanism is also similar to that of spheres, as each cube will be attracted to a crystal lattice node due to the local minimum potential energy there, and the ordered structure formed in the central region aligns with those formed at the face and edge boundaries, so the self-assembly follows a similar Markov chain network. Note in the experiments, we observed some cubic particles in the outermost layer may not settle at an ordered position first, but with the collision from the further added particles these defective particles would be changed to ordered. This also happens in the self-assembly of spheres occasionally. Therefore, the particles are not fully athermal, as each new particle will bring a small bit of energy to the system which can direct the new particle to be ordered and may also heal local defects. In this regard, this kind of self-assembly is near ground state.

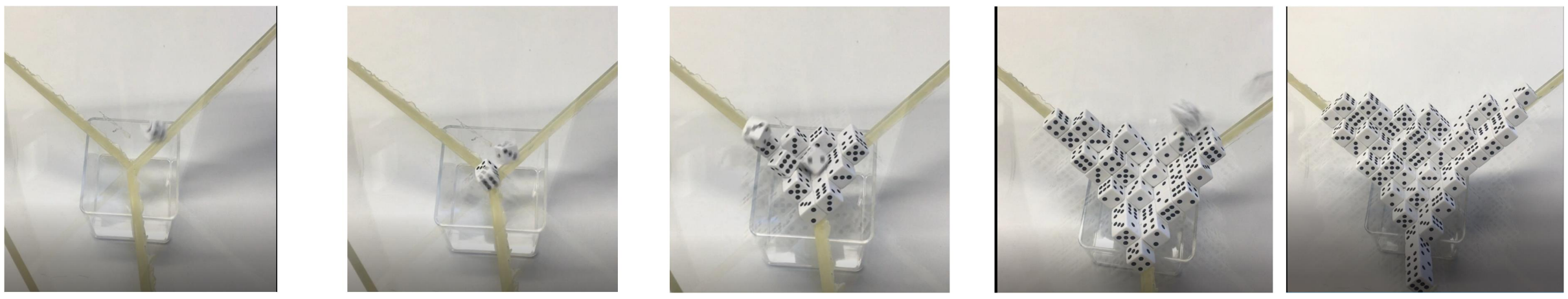

**FIG. 5.** Snapshots of a self-assembly process of mono-sized cubic dices in an inverted corner composed of three right triangles, top view. Movie:
https://drive.google.com/open?id=1lCNanDNjTfeJnFX2dbKvySWzfdHWDUax

The crystallization of randomly packed mono-sized spheres and cubes sheds new light on the self-assembly of particles and deserves further studies. Firstly, they are rather different from the random packings normally seen. It is argued that the random packing structure results from collective jamming. Yet here using RSA, collective jamming is prevented, but whether disordered or ordered structures will be formed is dominated by the container shape. This shows that disorder will also result from the conflict between the boundary and the internal structure. Secondly, the self-assembly process manifests the geometrical effect on the crystal structural dynamics found in different systems [41-43], especially the local jamming effect, which can help understand the self-assembly of the systems far from equilibrium and near ground state. Finally, the study reveals a general idea for directing self-assembly for macroscopic elements: to consider the synergistic effect of boundary and elements. As lacking "thermodynamics" means, self-assembly of these systems needs to be directed with bottom-up methods. Previous studies focused on tailoring the interactions between elements [1,3,44], whereas more recent studies suggested the control through boundaries [27,45,46]. Our study shows the two aspects should be considered synergistically to coordinate the order formed at boundaries and between elements. Here such coordination can critically transform a packing of mono-sized spheres from glass to crystal. This should be a general principle to be considered in directing self-assembly though how to achieve it can be different in other systems.

**Acknowledgments:** The authors are grateful for the financial support from Australian Research Council and Jiangsu Industrial Technology Research Institute (IH140100035) and the international PhD scholarships from Western Sydney University.

**Competing interests:** Authors declare no competing interests.

# Appendix

## 1 Experiments

We have randomly packed mono-sized spherical particles into crystal in the inverted half square pyramid with different particles. The tetrahedron container was made of plexiglass and simply sat on another container during the packing experiments, as shown in Fig. A1. Experiment movies can be seen on google drive:

- Ping-pong balls, size 40mm, pour packing: https://drive.google.com/open?id=1w0hgGUt9rJmveQfpQTx81EFXtqKGKbTS
- Tennis balls, size 66mm, one by one packing: https://drive.google.com/open?id=1g9FXHLhthH8GUp9q29TiESz40jH9ysZR
- Steel ball bearings, size 5mm, batch by batch packing: https://drive.google.com/open?id=1LQ-EMYZmi-rT0MfHKfLWotrqZwUjX34K

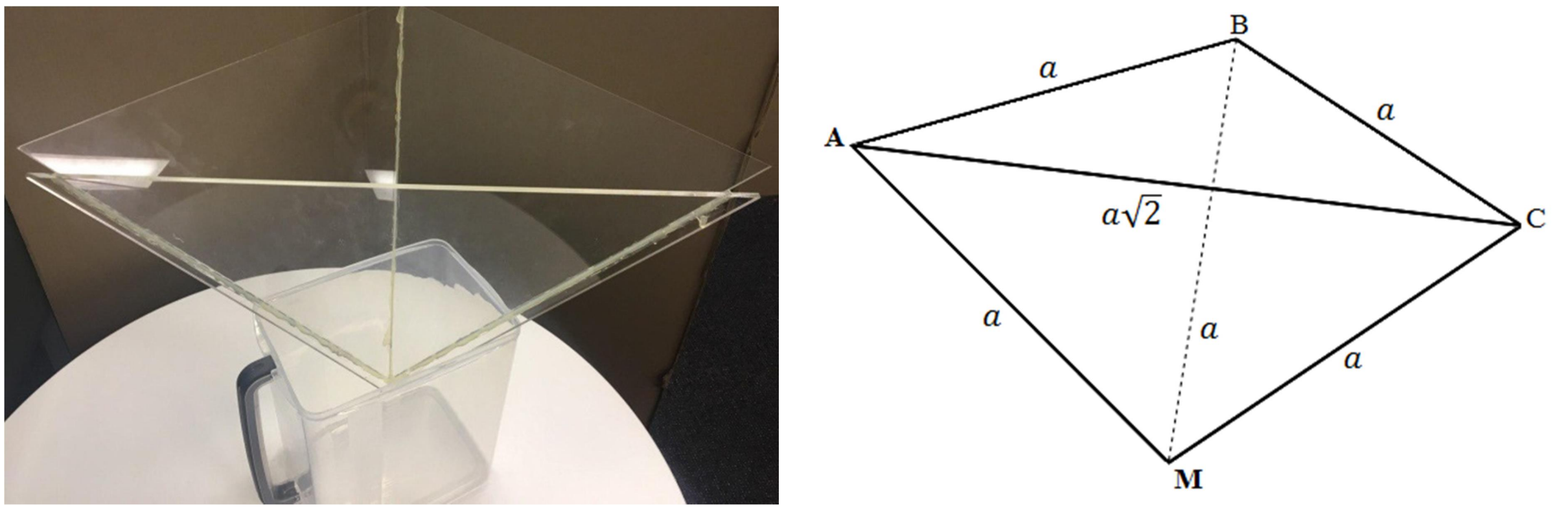

**(a)** **(b)**

**FIG. A1.** (a), The tetrahedron container used in experiments; and (b), the dimensions of the container ($a$=450mm).

## 2 Numerical simulations

The packing of mono-sized spheres in the HSP tetrahedron was also simulated by the discrete element method (DEM). Based on first principles, DEM can accurately predict behaviours of granular materials, which has been widely shown in the literature. The details of the model can be found in our recent studies [30,47]. The parameters used in the simulation are listed in Table A1, which have also been validated in previous studies [30,47]. Particles were fed batch by batch from the top of the container with their horizontal positions randomized. The number of particles per batch is denoted as $N_B$ and the interval time between two batches is denoted as $T_B$. The feed rate can be given by $N_r = N_B / T_B$. With the increase of $T_B$, the feeding changes from continuous to batch-by-batch.

**Table A1**. Parameters used in simulation.

| Parameter, *Symbol* (*Dimension*) | Value |
|---|---|
| Particle density, $\rho_p$ ($kg/m^3$) | 2500 |
| Particle diameter, $d$ ($mm$) | 5 |
| Young's modulus, $Y$ ($Pa$) | $10^7$ |
| Poisson ratio, $\sigma$ | 0.29 |
| Normal damping coefficient, $\gamma_n$ | 0.3 |
| Tangential damping coefficient, $\gamma_t$ | 0.3 |
| Sliding friction coefficient, $\mu_s$ | 0.3 |
| Rolling friction coefficient, $\mu_r$ | 0.001 |

Fig. A2a demonstrates the simulated packing of total 15000 particles with $N_B$=1 and $T_B$ = 0.5 secs. For the final packing, the local structure around each particle was analysed by the adaptive Common-Neighbour Analysis (a-CNA) method using the Ovito software [34]. Using this method, the particles as the centres of the local FCC clusters are coloured green and those in the clusters but not at the centres are coloured light green in Fig. A2a [30]. This method, however, is not applicable to particles along some container edges or in the incomplete top layers, as can be seen in Fig. A2a. Fig. A2b shows the fraction of the FCC particles to the total particles in the packing as a function of packing height, in which the particles contacting with the boundary are discarded. It can be seen that the FCC particles are always 100%. Comparing Fig. 1 and Fig. A2a, one can also notice that the particles on the *AMB* and *AMC* walls conform to {111} and {100} faces in the FCC crystal respectively, which is in accordance with our design.

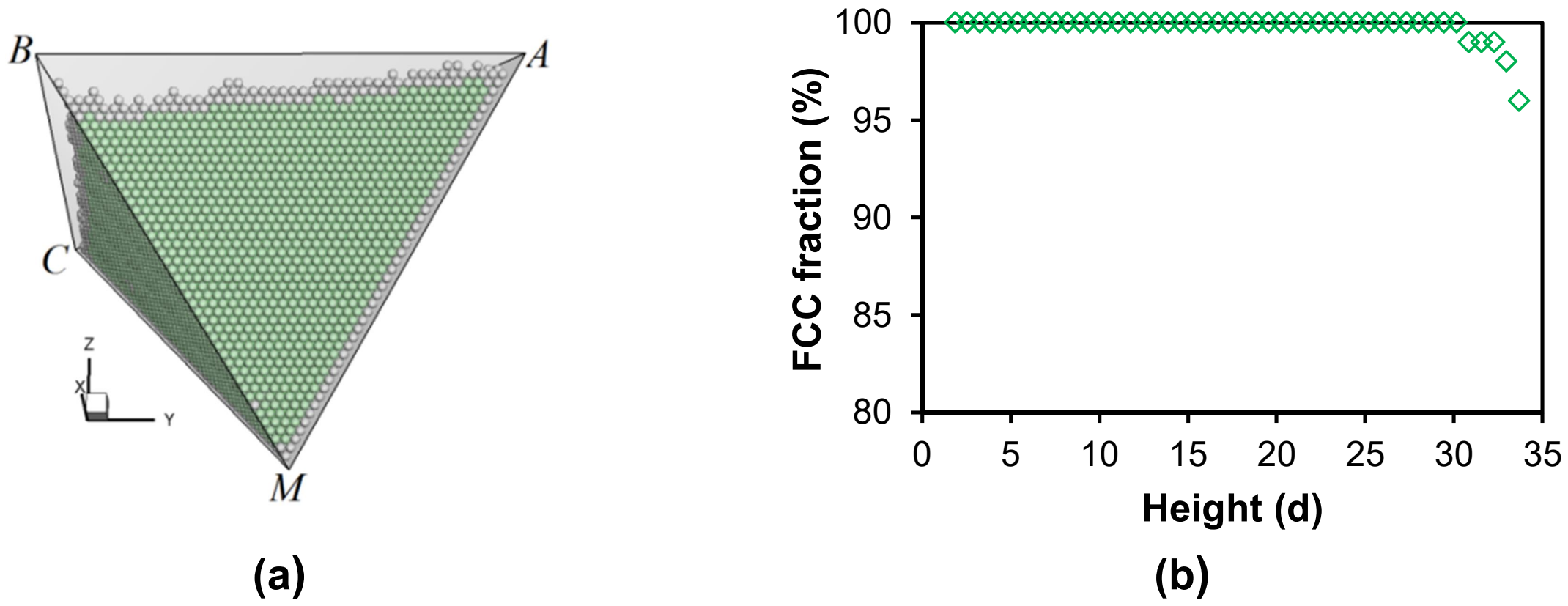

**FIG. A2.** (a), Simulated packing of 15000 particles in the HSP tetrahedron without vibration, particles were randomly dropped one by one; and (b), fraction of FCC particles in the packing from the bottom to different heights.

## 3 Rigorous proof of FCC

As shown in previous studies [30,47], the a-CNA method may identify the ordered local structures with a small distortion. To more rigorously check the structure, the packing was further characterized with the following two methods:

1. *Packing efficiency compared to the ideal FCC packing*. Though the packing fraction for the perfect FCC is $\frac{\pi}{3\sqrt{2}} \approx 0.740$, here we need to consider the effect of the boundary. For an ideal FCC packing in the inverted HSP tetrahedron, as schematically shown in Fig. A3a, particles are piled vertically in horizontal layers of square structures. It is not difficult to obtain the ideal FCC packing structure as follows.

The particle number in the $i^{\text{th}}$ horizontal layer, denoted as $k_i$, can be given by,

$$k_i = \frac{i(i+1)}{2} \tag{1}$$

Therefore the total number of particles, *N*, from layer 1 to $n$ is given by,

$$N(n) = \sum_{i=1}^{n} k_i = \frac{n(n+1)(n+2)}{6} \tag{2}$$

With the bottom of the inverted HSP tetrahedron at $z = 0$, the height of the particle centres in layer $n$ is given by,

$$H_c(n) = \frac{1+\sqrt{3}}{2}d + (n-1)\frac{\sqrt{2}}{2}d \tag{3}$$

where $\frac{1+\sqrt{3}}{2}d$ is the height of the first layer, and $\frac{\sqrt{2}}{2}d$ is the height of other layers (Fig. A3a). If counted from the top surface of the particles the height will be: $H(n) = H_c(n) + \frac{1}{2}d$. The volume of this container with a height $h$ is $V(h) = \frac{h^3}{3}$. So the ideal packing fraction as a function of $n$ is given by,

$$\rho_{ideal}(n) = \frac{N(n) \cdot V_p}{V(H(n))} = \pi \frac{n(n+1)(n+2)}{12\left(\frac{2+\sqrt{3}}{2} + (n-1)\frac{\sqrt{2}}{2}\right)^3} \tag{4}$$

where $V_p$ is the volume of a single particle. Note that $\lim_{n\to\infty} \rho_{ideal}(n) = \frac{\pi}{3\sqrt{2}}$, which matches the packing fraction of the FCC unit cell. We then define the packing efficiency as:

$$e(n) = \frac{\rho(n)}{\rho_{ideal}(n)} \tag{5}$$

where $\rho(n)$ is the packing fraction obtained in the real packing with the height to $H(n)$. In Fig. A3b, $e$ is plotted as a function of $n$ for the simulated packing, it can be seen that $e$ is literally 100%, except when the packing includes incomplete top layers.

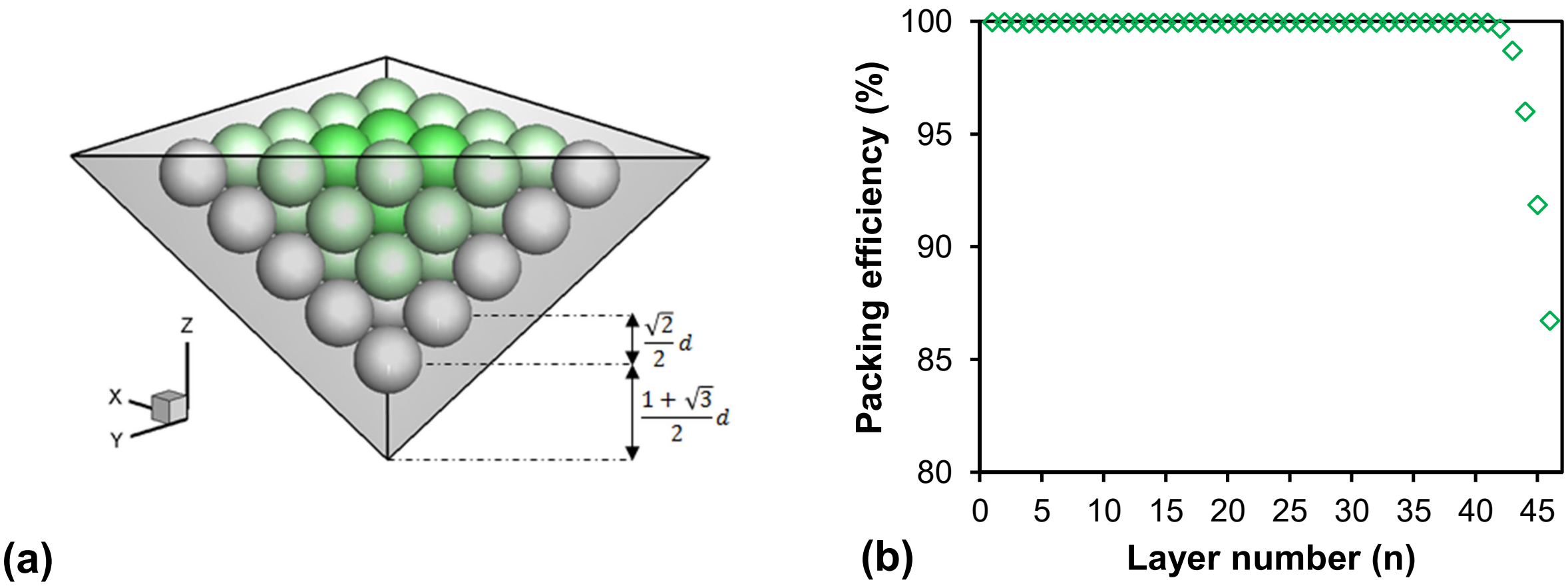

**FIG. A3.** (a), 5 layers of the FCC packing in the inverted HSP tetrahedron. (b), the distribution of the local packing fraction for the simulated packing shown in Fig. A2.

2. *Local packing fraction*. Voronoi analysis by Voro++ [48] was used to obtain the local packing fraction of each particle as the density of the Voronoi polyhedron enclosing it. For all the particles at the centre which are not contacting with the boundary or in the incomplete top layers, their local packing fractions were measured in the range of 0.740±0.005.

## 4 Reproducibility of the self-assembly

As shown in the experiments, this self-assembly can be achieved with different particles and at a small feed rate. Here we have also conducted simulations with different $N_B$ and repeated 9 simulations using different random seeds for each $N_B$ to test the reproducibility. Note 3000 particles were used for these simulations to save computational effort. With $N_B = 1$, we always obtain a perfect crystal. With $N_B > 1$, we can identify several empty lattice nodes without particles, as shown in Fig. A4a. They are rare point defects in the crystal. Fig. A4b shows the defect percentage calculated as the number of void nodes divided by the particle number. The defect percentage increased with increasing $N_{\mathrm{B}}$, whereas the percentage was less than 0.1% when $N_{\mathrm{B}}$=5. We observed a sharp increase of defect percentage for $N_{\mathrm{B}}$>20.

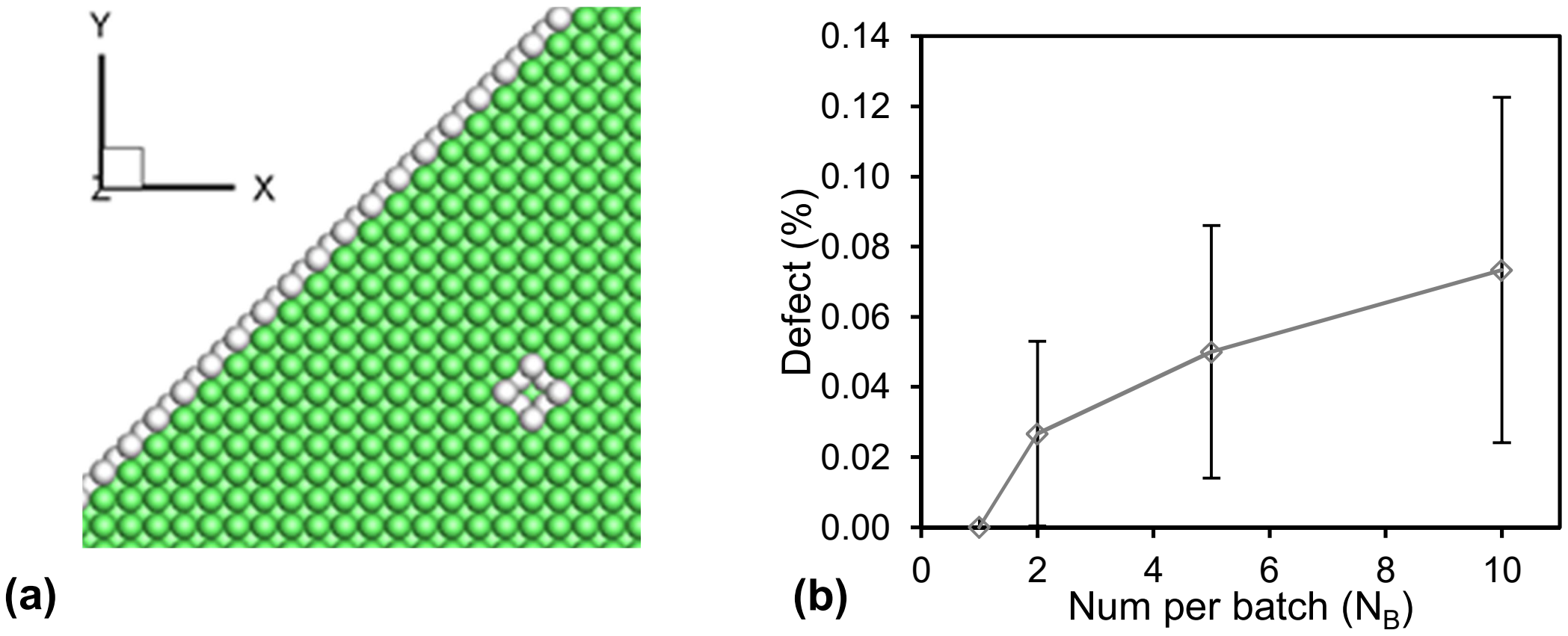

**FIG. A4.** (a), Point defect identified in a simulated packing with $N_{\mathrm{B}}$=5. (b), Defect percentage versus $N_B$. Nine simulations with different random seeds conducted for each $N_B$.

Moreover, we changed material properties in the simulation but can always obtain nearly perfect FCC packings, as shown in Fig. A5. Here we used $N_{\mathrm{B}}$ = 5 and $T_{\mathrm{B}}$= 0.5 secs (e.g., https://drive.google.com/open?id=1EBISx5PpEPYhN6lXCMFHz813uA8ZLZs8). For all these packings, we identified less than 0.2% defects.

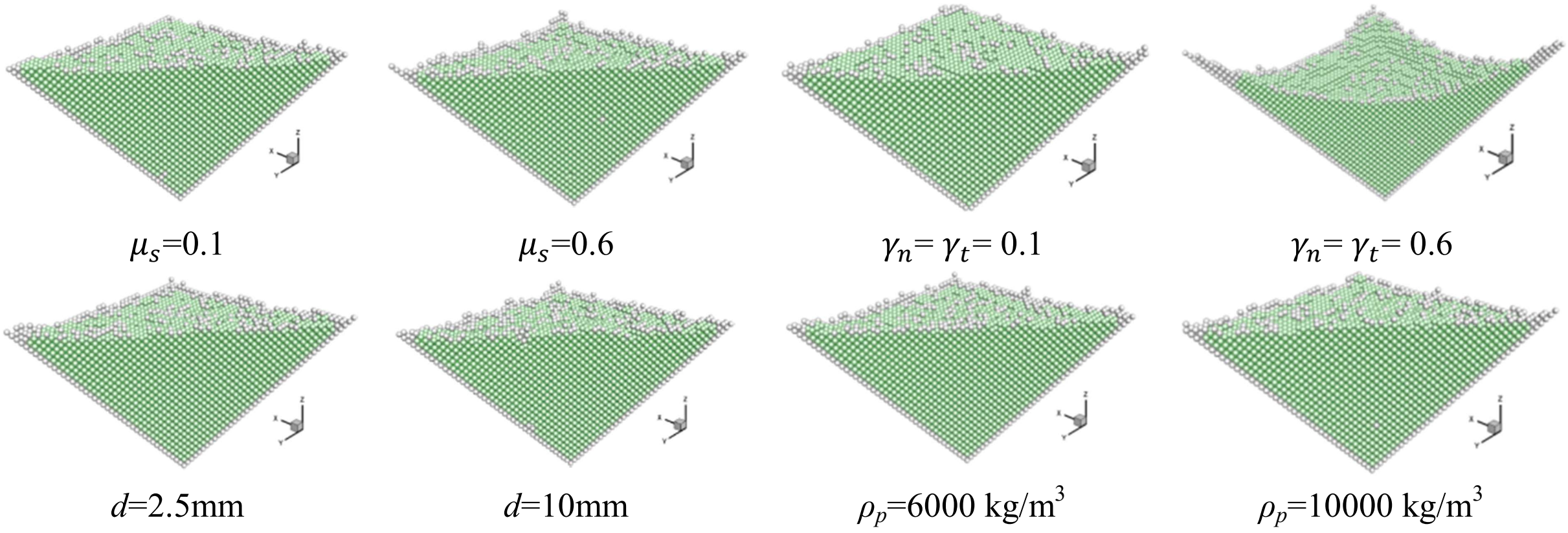

**FIG. A5.** Simulated self-assembly of mono-sized spheres with different material properties.

## 5 Energy analysis for packings in the inverted HSP tetrahedron

In the inverted HSP tetrahedron, the maximum and minimum gravitational energy for all possible packing states with a given particle number can be calculated. In order not to lose generality, we define the dimensionless gravitational energy $E_g$ as:

$$E_g = \sum_i^N \frac{mgz_i}{mgd} = \sum_i^N \frac{z_i}{d} \tag{6}$$

where $z_i$ is the $i^{\text{th}}$ particle's height. The maximum $E_g$ is for packings with particles just packed along an edge of the container in the JO1 structure, so the particle number $N$ equals to the layer number $n$. Then we have,

$$E_{g_{max}}(N) = \sum_{i=1}^{N} H_C(i) = \frac{\sqrt{2}}{4}N^2 + \left(\frac{2+2\sqrt{3}-\sqrt{2}}{4}\right)N \tag{7}$$

The minimum $E_g$, on the other hand, is for the ideal FCC packing with a smooth top layer or only the topmost layer is incomplete. For the former case, Eq. (2) gives $N(n) = \frac{(n)(n+1)(n+2)}{6}$, and $E_g$ can be given by:

$$E_{g_{min}}\big(N(n)\big) = n(n+1)\left[\frac{1+\sqrt{3}}{12}(n+2) + \frac{\sqrt{2}}{16}(n^2+n-2)\right] \tag{8}$$

If $N$ is between the values of the ideal packings with $n$ layers and $n+1$ layers, i.e., $N(n) < N < N(n+1)$, $E_{g_{min}}$ can be given by:

$$E_{g_{min}}(N) = E_{g_{min}}\big(N(n)\big) + [N - N(n)]\left(\frac{1+\sqrt{3}}{2} + \frac{\sqrt{2}}{2}n\right) \tag{9}$$

Eq. (7) gives the upper boundary and Eqs. (8) & (9) the lower boundary in Fig. 4.

## References

1 Damasceno, P. F., Engel, M. & Glotzer, S. C. Predictive Self-Assembly of Polyhedra into Complex Structures. *Science* **337**, 453-457, doi:10.1126/science.1220869 (2012).

2 Nagy, M., Akos, Z., Biro, D. & Vicsek, T. Hierarchical group dynamics in pigeon flocks. *Nature* **464**, 890-U899, doi:10.1038/nature08891 (2010).

3 Rubenstein, M., Cornejo, A. & Nagpal, R. Programmable self-assembly in a thousand-robot swarm. *Science* **345**, 795-799, doi:10.1126/science.1254295 (2014).

4 Bernal, J. D. A geometrical approach to the structure of liquids. *Nature* **183**, 141-147 (1959).

5 Jaeger, H. M., Nagel, S. R. & Behringer, R. P. Granular solids, liquids, and gases. *Reviews of Modern Physics* **68**, 1259-1273 (1996).

6 Liu, A. J. & Nagel, S. R. Nonlinear dynamics: Jamming is not just cool any more. *Nature* **396**, 21-22 (1998).

7 Torquato, S. & Stillinger, F. H. Jammed hard-particle packings: From Kepler to Bernal and beyond. *Reviews of Modern Physics* **82**, 2633-2672 (2010).

8 Torquato, S., Truskett, T. M. & Debenedetti, P. G. Is random close packing of spheres well defined? *Phys Rev Lett* **84**, 2064 (2000).

9 Behringer, R. P. & Chakraborty, B. The physics of jamming for granular materials: a review. *Reports on Progress in Physics* **82**, 012601, doi:10.1088/1361-6633/aadc3c (2018).

10 Baule, A., Morone, F., Herrmann, H. J. & Makse, H. A. Edwards statistical mechanics for jammed granular matter. *Reviews of Modern Physics* **90**, 015006 (2018).

11 Cieśla, M. & Kubala, P. Random sequential adsorption of cubes. *The Journal of chemical physics* **148**, 024501 (2018).

12 Wu, Y., An, X. & Yu, A. DEM simulation of cubical particle packing under mechanical vibration. *Powder technology* **314**, 89-101 (2017).

13 Malmir, H., Sahimi, M. & Tabar, M. R. R. Packing of nonoverlapping cubic particles: Computational algorithms and microstructural characteristics. *Physical Review E* **94**, 062901 (2016).

14 Asencio, K., Acevedo, M., Zuriguel, I. & Maza, D. Experimental Study of Ordering of Hard Cubes by Shearing. *Phys Rev Lett* **119**, 228002, doi:10.1103/PhysRevLett.119.228002 (2017).

15 Scott, G. & Kilgour, D. The density of random close packing of spheres. *Journal of Physics D: Applied Physics* **2**, 863 (1969).

16 O'Hern, C. S., Silbert, L. E., Liu, A. J. & Nagel, S. R. Jamming at zero temperature and zero applied stress: The epitome of disorder. *Physical Review E* **68**, doi:10.1103/Physreve.68.011306 (2003).

17 Kamien, R. D. & Liu, A. J. Why is Random Close Packing Reproducible? *Phys Rev Lett* **99**, 155501, doi:10.1103/PhysRevLett.99.155501 (2007).

18 Song, C., Wang, P. & Makse, H. A. A phase diagram for jammed matter. *Nature* **453**, 629-632, doi:10.1038/nature06981 (2008).

19 Hales, T. *et al.* in *Forum of Mathematics, Pi.* (Cambridge University Press).

20 Hales, T. C. A proof of the Kepler conjecture. *Annals of mathematics*, 1065-1185 (2005).

21 Panaitescu, A., Reddy, K. A. & Kudrolli, A. Nucleation and Crystal Growth in Sheared Granular Sphere Packings. *Phys Rev Lett* **108**, 108001 (2012).

22 Royer, J. R. & Chaikin, P. M. Precisely cyclic sand: Self-organization of periodically sheared frictional grains. *Proceedings of the National Academy of Sciences* **112**, 49-53 (2015).

23 Rietz, F., Radin, C., Swinney, H. L. & Schröter, M. Nucleation in Sheared Granular Matter. *Phys Rev Lett* **120**, 055701, doi:10.1103/PhysRevLett.120.055701 (2018).

24 Nahmad-Molinari, Y. & Ruiz-Suarez, J. Epitaxial growth of granular single crystals. *Phys Rev Lett* **89**, 264302 (2002).

25 Carvente, O. & Ruiz-Suárez, J. C. Self-assembling of dry and cohesive non-Brownian spheres. *Physical Review E* **78**, 011302 (2008).

26 Yu, A. B., An, X. Z., Zou, R. P., Yang, R. Y. & Kendall, K. Self-Assembly of Particles for Densest Packing by Mechanical Vibration. *Phys Rev Lett* **97**, 265501 (2006).

27 Morales-Barrera, D. A., Rodríguez-Gattorno, G. & Carvente, O. Reversible Self-Assembly (fcc-bct) Crystallization of Confined Granular Spheres via a Shear Dimensionality Mechanism. *Phys Rev Lett* **121**, 074302, doi:10.1103/PhysRevLett.121.074302 (2018).

28 Francois, N., Saadatfar, M., Cruikshank, R. & Sheppard, A. Geometrical Frustration in Amorphous and Partially Crystallized Packings of Spheres. *Phys Rev Lett* **111**, 148001 (2013).

29 Frenkel, D. Order through entropy. *Nature Materials* **14**, 9-12, doi:10.1038/nmat4178 (2015).

30 Amirifar, R., Dong, K., Zeng, Q. & An, X. Bimodal self-assembly of granular spheres under vertical vibration. *Soft Matter* **15**, 5933-5944, doi:10.1039/C9SM00657E (2019).

31 Bernal, J., Knight, K. & Cherry, I. Growth of crystals from random close packing. *Nature* **202**, 852-854 (1964).

32 Panaitescu, A. & Kudrolli, A. Epitaxial growth of ordered and disordered granular sphere packings. *Physical Review E* **90**, 032203 (2014).

33 Jensen, K., Pennachio, D., Recht, D., Weitz, D. & Spaepen, F. Rapid growth of large, defect-free colloidal crystals. *Soft Matter* **9**, 320-328 (2013).

34 Stukowski, A. Visualization and analysis of atomistic simulation data with OVITO–the Open Visualization Tool. *Modelling and Simulation in Materials Science and Engineering* **18**, 015012 (2009).

35 Lundrigan, S. E. & Saika-Voivod, I. Test of classical nucleation theory and mean first-passage time formalism on crystallization in the Lennard-Jones liquid. *The Journal of Chemical Physics* **131**, 104503 (2009).

36 Trudu, F., Donadio, D. & Parrinello, M. Freezing of a Lennard-Jones fluid: From nucleation to spinodal regime. *Phys Rev Lett* **97**, 105701 (2006).

37 Moroni, D., Ten Wolde, P. R. & Bolhuis, P. G. Interplay between structure and size in a critical crystal nucleus. *Phys Rev Lett* **94**, 235703 (2005).

38 Woodcock, L. V. Entropy difference between the face-centred cubic and hexagonal close-packed crystal structures. *Nature* **385**, 141-143 (1997).

39 An, X. Z. & Yu, A. B. Analysis of the forces in ordered FCC packings with different orientations. *Powder Technology* **248**, 121-130, doi:10.1016/j.powtec.2013.01.064 (2013).

40 Wu, Y., An, X., Qian, Q., Wang, L. & Yu, A. Dynamic modelling on the confined crystallization of mono-sized cubic particles under mechanical vibration. *Eur. Phys. J. E* **41**, 139 (2018).

41 Irvine, W. T., Bowick, M. J. & Chaikin, P. M. Fractionalization of interstitials in curved colloidal crystals. *Nature materials* **11**, 948 (2012).

42 Gómez, L. R., García, N. A., Vitelli, V., Lorenzana, J. & Vega, D. A. Phase nucleation in curved space. *Nat Commun* **6**, 6856 (2015).

43 Lewis, D. J., Zornberg, L. Z., Carter, D. J. & Macfarlane, R. J. Single-crystal Winterbottom constructions of nanoparticle superlattices. *Nature Materials*, 1-6 (2020).

44 Jiang, X. C., Zeng, Q. H., Chen, C. Y. & Yu, A. B. Self-assembly of particles: some thoughts and comments. *Journal of Materials Chemistry* **21**, 16797-16805, doi:10.1039/C1JM12213D (2011).

45 Schuppler, M., Keber, F. C., Kröger, M. & Bausch, A. R. Boundaries steer the contraction of active gels. *Nat Commun* **7**, 13120, doi:10.1038/ncomms13120 (2016).

46 Ross, T. D. *et al.* Controlling organization and forces in active matter through optically defined boundaries. *Nature* **572**, 224-229, doi:10.1038/s41586-019-1447-1 (2019).

47 Amirifar, R., Dong, K., Zeng, Q. & An, X. Self-assembly of granular spheres under one-dimensional vibration. *Soft Matter* **14**, 9856-9869, doi:10.1039/C8SM01763H (2018).

48 Rycroft, C. Voro++: a three-dimensional Voronoi cell library in C++. *Chaos* **19** (2009).